\documentclass[sn-mathphys]{sn-jnl}

\usepackage{physics}
\usepackage{caption}
\usepackage{tablefootnote}

\usepackage{multicol}
\usepackage{amsmath}
\usepackage{multirow}
\usepackage{amsfonts,amssymb}
\usepackage{tabularx}
\usepackage{multirow}
\usepackage{bbold}
\jyear{2021}%
\theoremstyle{thmstyleone}%

\theoremstyle{thmstyletwo}%
\usepackage{tabularx}
\usepackage{multirow}
\usepackage{wrapfig}
\usepackage{tabularx,multirow,array}
\usepackage{float}
\theoremstyle{thmstylethree}%

\begin{document}
\title[Boosted quantum and semi-quantum communication protocols]{Boosted quantum and semi-quantum communication protocols}
\author*[]{\fnm{Rajni} \sur{Bala}}\email{Rajni.Bala@physics.iitd.ac.in}
\author[]{\fnm{Sooryansh} \sur{Asthana}}\email{sooryansh.asthana@physics.iitd.ac.in}
\author[]{\fnm{ V.} \sur{Ravishankar}}\email{vravi@physics.iitd.ac.in}

\affil*[]{\orgdiv{Department of Physics}, \orgname{IIT Delhi}, \orgaddress{\street{Hauz Khas}, \city{New Delhi}, \postcode{110016}, \state{Delhi}, \country{India}}}

\abstract{ Secure quantum communication protocols based on prepare-and-measure scheme employ mutually unbiased bases. In these protocols, a large number of runs, in which different participants measure in different bases, simply go wasted.  In this paper, we show that it is possible to reduce the number of such runs by a suitable design of the key generation rule. This results in a  significant increase in the key generation rate (KGR). We illustrate this advantage by proposing   quantum key distribution protocols and semi-quantum key distribution protocols by employing effective qubits encoded in higher dimensional quantum systems. None of them demands  the preparation of entangled states as resources and a relatively large amount of information can be transferred. For this reason, we believe that our proposals are worth pursuing experimentally. 
}
\keywords{quantum network, quantum cryptography, quantum secure direct communication}

\maketitle


\abstract{

}
\keywords{quantum network, semi--quantum cryptography, layered quantum cryptography}

\maketitle
\section{Introduction}
\label{introduction}
The proposal of the seminal BB84 quantum key distribution protocol (QKDP) has resulted in the advent of the field of secure quantum communication \cite{Bennett84}. Since then,  the field has witnessed unprecedented growth \cite{gisin2002quantum,pirandola2020advances,chau2015quantum}. QKDPs employing quantum resources such as contextuality, entanglement, and non-orthogonality of states have been proposed \cite{Ekert91, Bennett92, Bennett93,bala2021contextuality}.

There are many proposals for entanglement-based QKDPs employing qubits and qudits. However, their realizations face many challenges, owing, in particular, to the low yield of entangled states ($\sim {\rm mHz}$) \cite{erhard2020advances}. In contrast, prepare-and-measure protocols do not suffer from these challenges, giving an edge over entanglement-based QKDPs   \cite{pirandola2020advances}. Yet another attractive family of protocols is semi-QKDPs (SQKDP)  as it requires only one participant to be quantum\footnote{ A quantum participant (QP) can prepare a state and measure it in any basis whereas a {\it classical} participant (CP) can prepare a state and perform a measurement only in the computational basis.}  \cite{iqbal2020semi,boyer2009semiquantum,xie2018semi,ye2018semi,pan2020semi,bala2022quantum,bala2022semi,zhou2019multi,gong2014continuous}. Thus, it mostly utilizes the existing classical resources and reduces the burden of availability of quantum resources, which makes SQKDPs appropriate in the current noisy-intermediate scale quantum (NISQ) regime of quantum communication \cite{yan2019semi,li2020new}.

To be of practical utility, however, either of the protocols should have a high communication rate. 
This has led to several proposals 
\cite{bradler2016finite, Inoue02, EfficientQKD_qubit, wang2010efficient}.
 However,  these protocols have one ubiquitous feature--they employ mutually unbiased bases. So, keys are generated only if measurements are performed in the same basis.  Consequently, the data of the rest of the rounds -- amounting to half of the total rounds or more-- is simply discarded,  thereby limiting the KGR.  SQKDPs also suffer from the same problem of having to discard data from many rounds. In fact, the number of rounds that do not contribute to key generation is even larger (growing up to $75\%$) \cite{Boyer07}. All in all, there is a stringent bar on the key generation rate.  We do note that when the prepare-and-measure protocol employs biased probabilities, only a small fraction of data is discarded. But then, it requires the preparation of states belonging to Fourier basis. Generation of such high-dimension states suffers from a decline in fidelity \cite{goel2022inverse}, which becomes severe as the dimension increases.

\begin{table}[!htb]
        \begin{minipage}{.5\linewidth}
      \centering
\resizebox{!}{1.6cm} {
\begin{tabular}{ | c| c| c |c|}
\hline
\multirow{3}{*}{}&\multirow{3}{3cm}{ \textbf{ QKD with mutually unbiased bases \cite{cerf2002security}}}&\multirow{3}{4cm}{\bf QKD based on orthogonal state encoding \cite{shu2022quantum}}&\multirow{3}{*}{\bf bQKDP} \\
 && &\\
&&&\\\hline
\multirow{3}{*}{\bf Quantum channel} &\multirow{3}{*}{\bf ideal}&\multirow{3}{*}{\bf  ideal}&\multirow{3}{*}{\bf ideal}\\
&&&\\
&&&\\\hline
\multirow{3}{*}{\bf Resource states} &\multirow{3}{2.5cm}{\bf qudits}&\multirow{3}{4cm}{\bf Two entangled and two separable states}&\multirow{3}{2cm}{\bf effective qubits}\\
&&&\\
&&&\\\hline
\multirow{3}{*}{\bf Yield of states }&\multirow{3}{*}{\textbf{$\sim kHz$}}&\multirow{3}{*}{{\bf $\sim mHz$}}&\multirow{3}{*}{ $\sim kHz$}\\
 &&&\\
&&&\\ \hline
\multirow{5}{3cm}{\bf security against eavesdropping}& \multirow{5}{*}{\bf non-orthogonality of states}&\multirow{5}{3cm}{\bf \textit{ additional} decoy states are required}& \multirow{5}{*}{\bf Non-orthogonality of states }\\
&&&\\
&&&\\
 &&&\\
 &&&\\
 \hline
 \multirow{3}{3cm}{\bf Quantum memory} &\multirow{3}{3cm}{\bf not required} & \multirow{3}{3cm}{\bf Required} &\multirow{3}{3cm}{\bf not required} \\
&&&\\
&&&\\\hline
\multirow{3}{3cm}{\bf $\#$ minimum bases used}&\multirow{3}{*}{$2$}&\multirow{3}{*}{ $1$}&\multirow{3}{*}{ $3$}\\
&&&\\ 
&&&\\ \hline
\multirow{3}{3cm}{{\bf Scope for generalisation with current technology }}&\multirow{3}{*}{ moderate*}&\multirow{3}{*}{{\bf difficult}}&\multirow{3}{*}{{\bf easy}}\\
&&&\\
&&&\\\hline
 \end{tabular}}
 \caption*{(a)}
    \end{minipage}%
    \begin{minipage}{.5\linewidth}
      \centering
\resizebox{!}{1.6cm}{
\begin{tabular}{ | c| c| c |c|}
\hline
\multirow{3}{*}{}&\multirow{3}{3cm}{ \textbf{Efficient mediated SQKD\cite{chen2021efficient}}}&\multirow{3}{2.5cm}{\bf Efficient SQKD \cite{pan2022semi} }&\multirow{3}{*}{\bf bSQKDP}\\
 && &\\
&&&\\\hline
\multirow{2}{*}{\bf Quantum channel} &\multirow{2}{*}{\bf ideal}&\multirow{2}{*}{\bf  ideal}&\multirow{2}{*}{\bf ideal}\\
&&&\\\hline
\multirow{2}{3cm}{\bf Third party} &\multirow{2}{2.5cm}{\bf required}&\multirow{2}{2cm}{\bf not required}&\multirow{2}{2cm}{\bf not required}\\
&&&\\\hline
 \multirow{3}{3cm}{\bf \# Classical participants}& \multirow{3}{*}{ $2$}& \multirow{3}{*}{ $1$}&\multirow{3}{*}{ $1$}\\
 &&&\\
 &&&\\
 \hline
\multirow{3}{3cm}{\bf Resource states} &\multirow{3}{2.5cm}{\bf Separable states}&\multirow{3}{2cm}{\bf bipartite entanglement}&\multirow{3}{2cm}{\bf effective qubits}\\
&&&\\
&&&\\\hline
\multirow{3}{3cm}{\bf Yield of states }&\multirow{3}{*}{ $\sim kHz$}&\multirow{3}{*}{{ $\sim mHz$}}&\multirow{3}{*}{ $\sim kHz$ }\\
 &&&\\
&&&\\ \hline
\multirow{3}{3cm}{\bf Fidelity of states }&\multirow{3}{*}{$\sim 90\%$ \cite{wang2019demand}}&\multirow{3}{*}{$\sim 90\%$ \cite{wang2019demand}}&\multirow{3}{*}{ $\sim 99\%$}\\
  &&&\\
&&&\\ 
&&&\\ \hline
\multirow{3}{3cm}{\bf Scope for generalisation to higher-dimensions}&\multirow{3}{*}{\bf  difficult}&\multirow{3}{*}{{\bf  difficult}}&\multirow{3}{*}{\bf easy}\\
 &&&\\
&&&\\ \hline 
\end{tabular}
}
\caption*{(b)}
    \end{minipage}
    \caption{Comparison of the  (a) bQKDP, and (b) bSQKDP proposed in this paper with existing protocols. The symbol $*$ represents the decrement in fidelity with increasing dimensions. }
      \label{comparison}
\end{table}

   In this work, we show that the problem associated with KGR can be overcome and that too in an experiment-friendly manner. For this, we choose such bases whose measurements in different bases also contribute to the generation of key symbols. For experimental feasibility, we identify the states which are not only prepared easily but also with high fidelity. Parenthetically, we note that  effective qubits encoded in qudits have been generated with fidelities of $\sim 99\%$ \cite{ding2017high}, thereby making QKDPs involving them  experimentally more feasible.

Taking high yield of qubits encoded in qudits into account, we propose two new protocols, which we shall designate as boosted QKD (bQKD) and boosted SQKD (bSQKD). We make use of the feature that the  effective qubits encoded in a high-dimensional space  may carry more than $1$ bit per transmission. This is in contrast to physical qubit systems belonging to a two-dimensional Hilbert space.  This naturally results in an increase in KGR.   As illustrations, we present bQKD and bSQKD protocols  employing ququarts (bQKDP$_4$/bSQKDP$_4$) and quhexes (bQKDP$_6$) (section (\ref{QKD}) and (\ref{boosted SQKD})). We show 
the robustness of the protocols by analyzing it against various eavesdropping strategies (section (\ref{security_QKD}) and (\ref{Robustness_SQKD})). Thereafter, we generalize the protocol to qudit systems (section (\ref{generalisation})).  For a quick comparison, in tables (\ref{comparison} {\color{blue}(a)}) and (\ref{comparison} {\color{blue}(b)}), we compare various features of our protocols with those of already existing ones. Section (\ref{conclusion}) concludes the paper.

\section{Boosted QKDP (bQKDP)}
\label{QKD}

In this section, we present a protocol that allows sharing of keys securely when the number of participants is two. We start with two examples- (i) effective qubits encoded in ququart systems and, (ii) effective qubits encoded in quhex systems. The two examples show how the key generation rate increases considerably with an increase in the dimensionality of the space.  After that, we generalize these protocols to qudit systems (with $d$ being even) and deduce the corresponding KGR.
\subsection{bQKDP with ququarts (bQKDP$_4$)}
\label{QKD_four}
Let Alice and Bob be the two participants who want to share a key by employing ququarts.\\

\noindent{\textit{\textbf{Aim:}}} distribution of a key with minimum discarding of data.\\
\noindent{\textit{\textbf{Resources:}}} Three bases sets $B_0, B_1, B_2$ as given below are employed,
\begin{align}
    B_0 &= \{\ket{0}, \ket{1}, \ket{2}, \ket{3}\};\nonumber\\
    B_1 &= \Big\{\frac{1}{\sqrt{2}}\Big(\ket{0}\pm\ket{1}\Big), \frac{1}{\sqrt{2}}\Big(\ket{2}\pm \ket{3}\Big)\Big\};~B_2 = \Big\{\frac{1}{\sqrt{2}}\Big(\ket{0}\pm \ket{3}\Big), \frac{1}{\sqrt{2}}\Big(\ket{1}\pm \ket{2}\Big)\Big\}.
    \label{basis_ququart}
\end{align}

\begin{center}
    \textbf{The protocol}
\end{center}

 \noindent The steps of the protocol are as follows:
 \begin{enumerate}
     \item  Alice prepares a state randomly from $B_0, B_1$ or $B_2$ with equal probability and sends it to Bob. 
\item Bob measures the received state in one of the bases $B_0$, $B_1$, or $B_2$ with  equal probability. 
    \item Steps (1-2) are repeated for a sufficiently large number of rounds. Thereafter, Alice and Bob  reveal the bases that have been employed in each round on an authenticated classical channel.
    \item {\it Detection of eavesdropping:}
     To check if there is eavesdropping,  Alice and Bob choose a subset of rounds and compare the corresponding data. In the absence of eavesdropping, there would be no errors. Otherwise,  the protocol is aborted.
     \item {\it Generation of key:}
      The rounds in which Alice and Bob have chosen bases either from the set $\{B_0, B_1\}$ or $\{B_0, B_2\}$ generate key letters. The data of the rest of the rounds (which are two in number) is discarded. 
 \end{enumerate}
 \noindent{\textbf{\textit{Key generation rule:}}}
\begin{enumerate}
    \item  Whenever  Alice and Bob choose the same bases, four key letters are generated, amounting to $2$ bits of information. 
    \item  In addition to this, when Alice chooses the basis $B_0$ and Bob measures either in the basis $B_1$ or $B_2$, two key letters are generated as has been shown explicitly in the table (\ref{tab:distinct_sooryansh}). The same rule exists if Alice sends states from either basis $B_1$ or $B_2$ and Bob measures in the basis $B_0$.
\end{enumerate} 
      \begin{minipage}{.55\linewidth} %
\includegraphics[width=6.3cm]{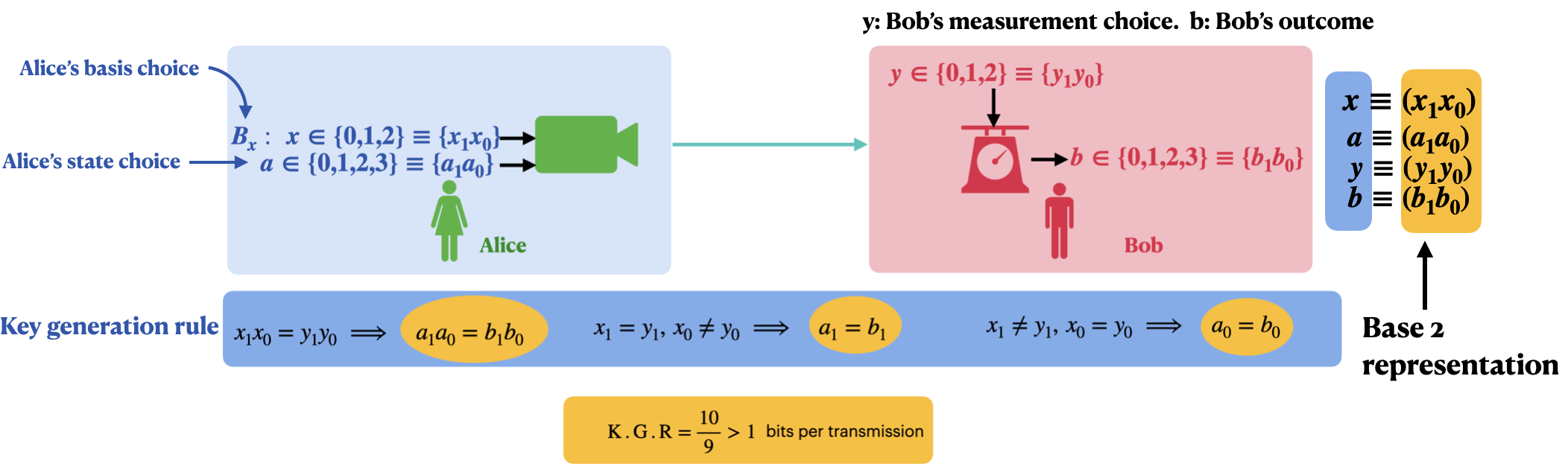}
\captionof{figure}{Pictorial representation of bQKDP$_4$. }
\label{fig:EQKD}
\end{minipage} %
\begin{minipage}{.4\linewidth} %
      \resizebox{!}{0.9cm}{
\begin{tabular}{|c|c|c|c|c|} 
 \hline
\multirow{3}{*}{{\bf Alice's basis}}& \multirow{3}{2cm}{{\bf Alice's state}}&\multirow{3}{*}{{\bf Bob's basis}}& \multirow{3}{2.5cm}{{\bf Post-measurement state of Bob}}&\multirow{3}{*}{{\bf Key letter}}\\
 &&&& \\
 &&&&\\ \hline\hline
\multirow{4}{*}{${\bf B_0}$} & \multirow{2}{*}{ ${\bf \ket{0}\slash\ket{1}}$}&\multirow{4}{*}{${\bf B_1}$}&\multirow{2}{*}{${\bf \frac{1}{\sqrt{2}}\big(\ket{0}\pm \ket{1}\big)}$}&\multirow{2}{*}{${\bf 0}$}\\
&&&&\\
&\multirow{2}{*}{${\bf \ket{2}\slash\ket{3}}$}& &\multirow{2}{*}{${\bf \frac{1}{\sqrt{2}}\big(\ket{2}\pm \ket{3}\big)}$}&\multirow{2}{*}{${\bf 1}$}\\
&&&&\\
  \hline
\multirow{4}{*}{${\bf B_0}$}  & \multirow{2}{*}{ ${\bf \ket{0}/\ket{3}}$}&\multirow{4}{*}{${\bf B_2}$}&\multirow{2}{*}{${\bf \frac{1}{\sqrt{2}}\big(\ket{0}\pm \ket{3}\big)}$}&\multirow{2}{*}{${\bf 0}$}\\
&&&&\\
&\multirow{2}{*}{${\bf \ket{1}/\ket{2}}$}&&\multirow{2}{*}{${\bf \frac{1}{\sqrt{2}}\big(\ket{1}\pm \ket{2}\big)}$}&\multirow{2}{*}{${\bf 1}$}\\
&&&&\\
  \hline
\end{tabular}
}
\captionof{table}{Key generation rule between Alice and Bob when both measure in different bases.}
        \label{tab:distinct_sooryansh}
\end{minipage}
   

The description of bQKDP$_4$  
is shown explicitly in figure (\ref{fig:EQKD}). 

\subsubsection{KGR}  
\label{KGR_QKD} 
 
In the protocol, both Alice and Bob choose one of three bases with equal probability of $1/3$. Additionally, Alice sends states with equal probability. With this information and following the key generation rule, $2$ bits of information are generated whenever Alice and Bob choose the same basis. However, in other cases, $1$ bit of information is generated. Thus, the KGR of the  bQKDP$_4$ is:
\begin{equation}
    r=\frac{1}{3^2}\big(2\times 3+1\times 4\big)=\frac{10}{9} {\rm ~bits/transmission.}
\end{equation}
Clearly, there is an increase in KGR as compared to the BB84 protocol with a ququart. The KGR of the latter is 1 bit per transmission.

  In the above analysis, we have considered unbiased probabilities at both Alice's and Bob's end in choosing a particular operation (states in the case of Alice and measurements in the case of Bob). However, one can also choose different bases with biased probabilities, introduced in \cite{ardehali1998efficient}, which will further enhance the KGR.


\subsection{Security of the bQKDP$_4$}
\label{security_QKD}
In this section, analogous to the security analyses given in \cite{yin2013three, sun2016multi,  zhu2018semi, tsai2018semi},  we show the robustness of the protocol against various eavesdropping strategies. The aim of Eve is to obtain information about the key being shared without getting detected. However, Eve will not be able to do so, thanks to the non-orthogonality of states. We start with a subspace attack that Eve may employ. This attack can be implemented with current technology.
\subsubsection{Subspace attack} In the bQKDP$_4$, key symbols are shared even if Alice and Bob choose different bases. Due to this, Eve may design a strategy that allows her to gain partial information with minimal errors, if not without errors. So, Eve may perform a subspace attack in which she measures an observable, say, $O$ defined as:
\begin{equation}    O\equiv\lambda_1\big(\ket{0}\bra{0}+\ket{1}\bra{1}\big)+\lambda_2\big(\ket{2}\bra{2}+\ket{3}\bra{3}\big),~~~~~~~\lambda_1\neq \lambda_2.
\end{equation}
In such an attack, Eve gets information of $1$ bit without introducing any errors, whenever Alice or Bob or both choose the bases $B_0$ or $B_1$. However, in the rounds in which both Alice and Bob choose the basis $B_2$, Eve introduces errors with a probability of $0.5$. So, in $l$ such rounds, Eve gets detected with a probability $(1-0.5^l)$.
Let $1- \epsilon$ be the desired probability of Eve’s detection, then,
 $l =o\Big(\log \frac{1}{\epsilon}\Big)$, which is a weak dependence on $\epsilon^{-1}$.


\subsubsection{Intercept-resend attack}
In such attacks, Eve measures intercepted states in the bases used for key generation. In the bQKDP$_4$, the three bases, {\it viz.}, $B_0, B_1, B_2$ are employed for key generation. So, Eve may employ all three bases to obtain information about a key or one of the bases. 
We discuss these possibilities in detail.\\

\noindent\textit{Strategy I - Eve measures in all three bases randomly:} This strategy of Eve corresponds to the standard case which is considered in all prepare-and-measure protocols. 
\begin{itemize}
\item In the rounds in which Eve's basis choice matches with either of Alice or Bob or  with both, she obtains full information being shared without introducing any errors. 
\item The rounds in which her basis choice does not match with either of Alice or Bob, her interventions introduce errors and hence get detected. The details for the same are given in table (\ref{tab:Eve_QKD} {\color{blue}(a)}).
\end{itemize}
\begin{table}[!htb]
    \begin{minipage}{.5\linewidth}
\resizebox{!}{0.68cm}{
\begin{tabular}{|c|c|c|c|c|} 
 \hline
\multirow{5}{2cm}{ {\bf Alice's and Bob's basis}}& \multirow{5}{2cm}{{\bf Eve's basis}}&\multirow{5}{2cm}{{\bf Information gain of Eve} }&\multirow{5}{2.35cm}{{\bf Probability of detecting Eve (in one round)}}&\multirow{5}{2.35cm}{{\bf Probability of detecting Eve (in $l$ rounds)}}\\
  & & &&\\
  &&&&\\
  &&&&\\
  &&&&\\
     \hline\hline
${\bf {B_0(B_1)}}$  & ${\bf B_1(B_0)}$&${\bf 1}$ {\bf bit}&${\bf 0.5}$&${\bf 1-0.5^l}$\\\hline
${\bf B_0(B_2)}$  & ${\bf B_2(B_0)}$&${\bf 1}$ {\bf bit}&${\bf 0.5}$&${\bf 1-0.5^l}$\\\hline
${\bf B_1(B_2)}$  & ${\bf B_2(B_1)}$&${\bf 0}$ {\bf bit}&${\bf 0.75}$&${\bf 1-0.25^l}$\\\hline
\end{tabular}}
 \caption*{(a)}
    \end{minipage}%
    \begin{minipage}{.5\linewidth}
\begin{center}
 \resizebox{!}{0.75cm}{
\begin{tabular}{|c|c|c|c|c|} 
 \hline
\multirow{4}{*}{\bf  Alice's basis} & \multirow{4}{*}{\bf Bob's basis}&\multirow{4}{2cm}{\bf Eve's information gain }&\multirow{4}{2.35cm}{\bf Probability of detecting Eve (in one round)}&\multirow{4}{2.35cm}{\bf Probability of detecting Eve (in $l$ rounds)}\\
&&& &\\
&&&&\\
 &&& &\\
     \hline\hline
${\bf B_0}$&${\bf B_0}$  &${\bf 1}$ {\bf bit}  &${\bf 0.5}$&${\bf 1-0.5^l}$\\\hline
${\bf B_0}$&${\bf B_2}$  &${\bf 0}$ {\bf bit} &${\bf 0.5}$&${\bf 1-0.5^l}$\\\hline
${\bf B_2}$&${\bf B_0}$  &${\bf 0}$ {\bf bit}&${\bf 0.5}$&${\bf 1-0.5^l}$\\\hline
${\bf B_2}$&${\bf B_2}$ &${\bf 0}$ {\bf bit}&${\bf 0.75}$&${\bf 1-0.25^l}$
\\\hline
\end{tabular}}
\caption*{(b)}
    \end{center}
    \end{minipage} 
     \caption{Eve's information gain and probability of detection of eavesdropping (a) for Eve's strategy I (in which she measures in all three bases randomly,  (b) for Eve's strategy III (when she measures in the basis $B_1$).}
      \label{tab:Eve_QKD}
\end{table}

\noindent{\it Strategy II - Eve measures only in basis $\boldsymbol{B_0}$:} In this case, whenever both Alice and Bob measure in the basis $B_0$, Eve gets complete information ($2$ bits) without introducing any errors.  However, in the rounds in which both Alice and Bob choose either bases $B_1$ or $B_2$, Eve gets partial information ($1$ bit) and also introduces errors with a probability of $0.5$. Thus, Eve's interventions, in $l$ such rounds, get detected with a $(1-0.5^l)$ probability which approaches unity for a sufficiently large $l$.\\

\noindent{\textit{Strategy III - Eve measures in basis $\boldsymbol{B_1}$:}} In such an attack, whenever either Alice or Bob or both measure in the basis $B_1$, Eve gets full information being shared without introducing any error. In the rest of the cases, Eve's interventions introduce errors and hence she gets detected.  This has been detailed in the table (\ref{tab:Eve_QKD} {\color{blue}(b)}). A similar analysis holds if Eve measures in the basis $B_2$.

From the above discussion, it is clear that strategy II is more advantageous from Eve's point of view as it provides her with more information and introduces errors only in two rounds.

\subsubsection{Entangle-and-measure attack}
In this attack, Eve entangles her ancilla with a traveling state. The effect of this operation can be expressed as:
\begin{align}    &\ket{i}_T\ket{0}_E\xrightarrow{U}\ket{ii}_{TE}, ~~~~~\forall i\in \{0,1,2,3\}.
\end{align}
 As is clear from the above equation, when Alice employs the basis $B_0$ Eve's gains information about the key without introducing any errors.
However, this will not be the case whenever the rest of the two bases, {\it viz.}, $B_1$ and $B_2$ are employed. This can be understood by  the effect of Eve's entangling operations on the states belonging to $B_1$ and $B_2$ as follows:
\begin{align}
    \dfrac{1}{\sqrt{2}}\Big(\ket{i}\pm\ket{j}\Big)\xrightarrow{U}\dfrac{1}{\sqrt{2}}\Big(\ket{ii}\pm\ket{jj}\Big),~ \ket{i}\pm \ket{j}\in B_1,~ B_2.
\end{align}
So, Eve's interventions are detected with a probability of $0.5$ whenever Bob uses the same bases as those of Alice. In $l$ such rounds, Eve gets detected with a probability of $1-(0.5)^l$ which approaches unity for a  sufficiently large $l$.

We next present the robustness of the bQKDP$_4$ against a general entangling attack, in which Eve interacts with an arbitrary unitary operation.
\subsubsection{General entangling attack}
The effect of this general entangling attack is represented with a unitary $U$ whose action in the computational basis can be expressed as:
\begin{align}    \label{eq:GEA}\ket{i}_T\ket{0}_E\xrightarrow{U}\sum_{j=0}^3\ket{j}_T\ket{E_{ij}}_E,
\end{align}
{\it where  Eve's ancilla states, viz., $\ket{E_{ij}}$ are not necessarily of the unit norm, or, for that matter, orthogonal}.\\ In the protocol, key letters are generated not only when Alice and Bob's basis choice matches but also even if their basis choices differ. Therefore, Eve's interceptions can be detected even if Alice and Bob choose different basis. We discuss these two cases separately.
\subsubsection*{(I) Alice and Bob choose the same basis}
Since in the protocol, Alice employs three bases, in the following, we show that Eve's tampering gets detected with a finite probability in all three cases. \\

\noindent{\textbf{\textit{(a) Alice and Bob choose basis $\boldsymbol{B_0}$:}}} Whenever Alice sends a state $\ket{i}\in B_0$, Bob gets the correct result, following equation (\ref{eq:GEA}), only with a probability $\norm{\ket{E_{ii}}}^2< 1$. Thus, Alice and Bob can detect the presence of Eve by comparing a subset of rounds with a probability $(1-\norm{\ket{E_{ii}}}^2)$. In $l$ such rounds, the presence of Eve gets detected with a probability $\big(1-\norm{\ket{E_{ii}}}^{2l}\big)$.\\ 
\noindent{\textbf{\textit{(b) Alice and Bob choose basis $\boldsymbol{B_1}$:}}} The rounds in which Alice sends a state $\ket{i}\pm\ket{j}$, Bob gets the correct result with probability, $p_\pm= \norm{\frac{1}{2}\big(\ket{E_{ii}}\pm\ket{E_{ij}}\pm\ket{E_{ji}}+\ket{E_{jj}}\big)}^2$. Thus, Eve's tampering gets detected with a probability $(1-p_\pm)$.\\
 A similar analysis holds when Alice and Bob choose the basis $B_2$. The difference will only be in the values of $i$ and $j$  for states $\ket{i}\pm\ket{j}$.\\A detailed analysis is given in table (\ref{tab:Comparison_QKD_123} {\color{blue}(a)}).  We now move on to the case when Alice and Bob choose different basis.


\begin{table}[!htb]
        \begin{minipage}{.6\linewidth}
\resizebox{6.75cm}{!}{
    \centering
\begin{tabular}{|c | c| c|c|}
\hline
\multirow{3}{3cm}{\bf\centering Alice and Bob's basis choice} &\multirow{3}{3cm}{\bf State sent by Alice}&\multirow{3}{4cm}{\bf Probability of getting correct outcome}&\multirow{3}{3cm}{\bf Probability of detecting Eve}\\
 &&& \\
&&&\\\hline
\multirow{4}{*}{\large $\bf{B_0}$}&\multirow{4}{*}{\large$\bf\ket{i}$} &\multirow{4}{*}{\large$\bf\norm{\ket{E_{ii}}}^2$}&\multirow{4}{*}{\large$\bf 1-\norm{\ket{E_{ii}}}^2$}\\
&&&\\
&&&\\&&&\\\hline
\multirow{12}{*}{\large$\bf B_1$}&\multirow{3}{*}{\large$\bf \ket{0}+\ket{1}$} &\multirow{3}{*}{\large$\bf \frac{1}{4}\norm{\sum_{i,j=0}^1\ket{E_{ij}}}^2$}&\multirow{3}{*}{\large$\bf 1-\frac{1}{4}\norm{\sum_{i,j=0}^1\ket{E_{ij}}}^2$}\\
&&&\\
&&&\\
&\multirow{3}{*}{\large$\bf\ket{0}-\ket{1}$} &\multirow{3}{*}{\large$\bf\frac{1}{4}\norm{\sum_{i,j=0}^1(-1)^{i+j}\ket{E_{ij}}}^2$}&\multirow{3}{*}{\large$\bf 1-\frac{1}{4}\norm{\sum_{i,j=0}^1(-1)^{i+j}\ket{E_{ij}}}^2$}\\
&&&\\
&&&\\
&\multirow{3}{*}{\large$\bf \ket{2}+\ket{3}$} &\multirow{3}{*}{\large$\bf \frac{1}{4}\norm{\sum_{i,j=2}^3\ket{E_{ij}}}^2$}&\multirow{3}{*}{\large$\bf 1-\frac{1}{4}\norm{\sum_{i,j=2}^3\ket{E_{ij}}}^2$}\\
&&&\\
&&&\\
&\multirow{3}{*}{\large$\bf \ket{2}-\ket{3}$} &\multirow{3}{*}{\large$\bf \frac{1}{4}\norm{\sum_{i,j=2}^3(-1)^{i+j}\ket{E_{ij}}}^2$}&\multirow{3}{*}{\large$\bf 1-\frac{1}{4}\norm{\sum_{i,j=2}^3(-1)^{i+j}\ket{E_{ij}}}^2$}\\
&&&\\
&&&\\\hline
\multirow{12}{*}{\large$\bf B_2$}&\multirow{3}{*}{\large$\bf \ket{0}+\ket{3}$} &\multirow{3}{*}{\large$\bf \frac{1}{4}\norm{\sum_{i,j=0,3}\ket{E_{ij}}}^2$}&\multirow{3}{*}{\large$\bf 1-\frac{1}{4}\norm{\sum_{i,j=0,3}^1\ket{E_{ij}}}^2$}\\
&&&\\
&&&\\
&\multirow{3}{*}{\large$\bf \ket{0}-\ket{3}$} &\multirow{3}{*}{\large$\bf \frac{1}{4}\norm{\sum_{i,j=0,3}(-1)^{i+j}\ket{E_{ij}}}^2$}&\multirow{3}{*}{\large$\bf 1-\frac{1}{4}\norm{\sum_{i,j=0,3}(-1)^{i+j}\ket{E_{ij}}}^2$}\\
&&&\\
&&&\\
&\multirow{3}{*}{\large$\bf \ket{1}+\ket{2}$} &\multirow{3}{*}{\large$\bf \frac{1}{4}\norm{\sum_{i,j=1}^2\ket{E_{ij}}}^2$}&\multirow{3}{*}{\large$\bf 1-\frac{1}{4}\norm{\sum_{i,j=1}^2\ket{E_{ij}}}^2$}\\
&&&\\
&&&\\
&\multirow{3}{*}{\large$\bf \ket{1}-\ket{2}$} &\multirow{3}{*}{\large$\bf \frac{1}{4}\norm{\sum_{i,j=1}^2(-1)^{i+j}\ket{E_{ij}}}^2$}&\multirow{3}{*}{\large$\bf 1-\frac{1}{4}\norm{\sum_{i,j=1}^2(-1)^{i+j}\ket{E_{ij}}}^2$}\\
&&&\\
&&&\\\hline
  \end{tabular}}
   \caption*{(a) }
    \end{minipage}%
    \begin{minipage}{.4\linewidth}
 \resizebox{!}{1cm}{
    \centering
\begin{tabular}{|c |c| c|c| c|c|}
\hline
\multirow{3}{2cm}{\centering\bf Alice's basis choice} &\multirow{3}{2cm}{\bf State sent by Alice}&\multirow{3}{2cm}{\bf \centering Bob's basis choice} &\multirow{3}{4cm}{\bf Post-measurement state corresponding to correct outcome}&\multirow{3}{2.5cm}{\bf Probability of getting correct outcome}&\multirow{3}{2.5cm}{\bf Probability of detecting Eve}\\
 &&&&& \\
&&&&&\\\hline\hline
\multirow{6}{*}{\large$\bf B_0$}&\multirow{6}{*}{\large$\bf \ket{i}$} &\multirow{6}{*}{\large$\bf B_1/B_2$}&\multirow{3}{*}{\large$\bf \ket{i}+\ket{k}$}&\multirow{3}{*}{\large$\bf p_{i+}=\frac{1}{2}\norm{\ket{E_{ii}}+\ket{E_{ik}}}^2$}&\multirow{6}{*}{\large$\bf 1-p_{i+}-p_{i-}$}\\
&&&&&\\
&&&&&\\
& &&\multirow{3}{*}{\large$\bf \ket{i}-\ket{k}$}&\multirow{3}{*}{\large$\bf p_{i-}=\frac{1}{2}\norm{\ket{E_{ii}}-\ket{E_{ik}}}^2$}&\\
&&&&&\\
&&&&&\\\hline
\multirow{12}{*}{\large$\bf B_1/B_2$}&\multirow{6}{*}{\large$\bf \ket{i}+\ket{j}$} &\multirow{12}{*}{\large$\bf B_0$}&\multirow{3}{*}{\large$\bf \ket{i}$}&\multirow{3}{*}{\large$\bf p_{+i}=\frac{1}{2}\norm{\ket{E_{ii}}+\ket{E_{ji}}}^2$}&\multirow{6}{*}{\large$\bf 1-p_{+i}-p_{+j}$}\\
&&&&&\\
&&&&&\\
&&&\multirow{3}{*}{\large$\bf \ket{j}$}&\multirow{3}{*}{\large$\bf p_{+j}=\frac{1}{2}\norm{\ket{E_{ij}}+\ket{E_{jj}}}^2$}&\\
&&&&&\\
&&&&&\\
&\multirow{6}{*}{\large$\bf \ket{i}-\ket{j}$} &&\multirow{3}{*}{\large$\bf \ket{i}$}&\multirow{3}{*}{\large$\bf p_{-i}=\frac{1}{2}\norm{\ket{E_{ii}}-\ket{E_{ji}}}^2$}&\multirow{6}{*}{\large$\bf 1-p_{-i}-p_{-j}$}\\
&&&&&\\
&&&&&\\
&&&\multirow{3}{*}{\large$\bf \ket{j}$}&\multirow{3}{*}{\large$\bf p_{-j}=\frac{1}{2}\norm{\ket{E_{ij}}-\ket{E_{jj}}}^2$}&\\
&&&&&\\
&&&&&\\\hline
  \end{tabular}
  }
    \caption*{(b) }
    \end{minipage} 
    \caption{Probability of getting the correct outcomes and Eve's detection (a) when both Alice and Bob choose the same basis, and (b) when Alice and Bob choose a different basis.}
            \label{tab:Comparison_QKD_123}
\end{table}
\subsubsection*{(II) Alice and Bob choose different basis}
For the rounds in which Alice and Bob choose different basis  either from the set $\{B_0, B_1\}$ or $\{B_0, B_2\}$, Eve's presence may be detected by comparing a subset of rounds.

\noindent{\textbf{\textit{(a) Alice and Bob choose basis $B_0$ and $B_1$:}}} Whenever Alice sends a state $\ket{i}\in B_0$, and Bob measures in the basis $B_1$, he gets the post measurement states $\ket{i}\pm \ket{j}$ with equal probability. However, due to Eve's interaction, Bob gets states $\ket{i}\pm \ket{j}$ with probability $p_{i\pm}=\norm{\frac{1}{\sqrt{2}}\big(\ket{E_{ii}}\pm \ket{E_{ij}}\big)}^2$. As a result, Eve gets detected with a probability $(1-p_{i+}-p_{i-})$.\\
A similar analysis holds when Alice chooses the basis $B_0$ and Bob chooses the basis $B_2$.

\noindent{\textbf{\textit{(b) Alice and Bob choose basis $B_1$ and $B_0$:}}} In this case, transmission of state $\ket{i}\pm\ket{j}$ by Alice provides Bob with post-measurement states $\ket{i}$ or $\ket{j} $ with equal probability in the absence of Eve.  However, due to Eve's interaction, Bob gets states $\ket{i}$ and $\ket{j}$ with probabilities $p_{\pm i}=\norm{\frac{1}{\sqrt{2}}\big(\ket{E_{ii}}\pm\ket{E_{ji}}\big)}^2$ and $p_{\pm j}=\norm{\frac{1}{\sqrt{2}}\big( \ket{E_{ij}}\pm\ket{E_{jj}}\big)}^2$. Thus, Eve's tampering gets detected with a probability $(1-p_{\pm i}-p_{\pm j})$.

A similar analysis holds for the other choice \textit{viz.,} $\{B_0,B_2\}$.
Details of possible cases are given in table (\ref{tab:Comparison_QKD_123} {\color{blue}(b)}).

This concludes our discussion on the robustness of the protocol. We, next, discuss the robustness of the protocol against implementation-based attacks.

\subsubsection{Photon-number-splitting (PNS) attack}
\label{PNS} 
Though theoretical proposals of QKDPs employ single photons, in experiments, these are implemented with weak coherent pulses (WCP). Employment of WCP facilitates Eve with opportunities that are otherwise unavailable. PNS attack is one such strategy. In this attack, Eve can take one photon off a multi-photon pulse and block all the single-photon pulses. This technique will help Eve in obtaining all the shared information without introducing any errors. Alice and Bob can detect this attack by employing an additional decoy pulse with a lower mean photon number as has been discussed in \cite{acin2004coherent}. Since both pulses are sent randomly, Eve cannot distinguish between the two. As a result, her interventions will change the statistics revealing her presence.

 \subsubsection{Trojan-horse attack}\label{Trojan}
 
 This attack is another example in which Eve employs imperfections of the devices employed to implement protocols, which was first introduced in \cite{gisin2006trojan}. One attack in this category is the `invisible photon attack'. In this attack, Eve sends photons of  different wavelengths to determine  the measurement settings of Bob's measurement device. These attacks can be circumvented by using a filter of a specified wavelength. As a result, only signal photons can reach the detector, thus removing the possibility of such attacks.
 
 Another important attack in this category is the `delay photon attack'. Bob  can detect this attack by counting the number of photons for a randomly chosen subset of the signal. This completes the robustness to eavesdropping attacks.

  This concludes our discussion of bQKDPs with ququart systems. In this protocol, data of the rounds in which Alice and Bob choose $B_1$ and $B_2$ respectively, and vice-versa are completely discarded. However, by employing effective qubits encoded in quhex  and, for that matter, in qudit ($d\geq 6)$ systems, even some of these runs can be employed for key generation. This further limits the data which is to be discarded and so using resources in their full form. We show it explicitly by presenting a bQKDP with quhex systems designated as bQKDP$_6$ in Appendix (\ref{QKD_six}).

  In the subsequent section, we generalize the protocol to qudit systems, with $d$ being even.

\section{Generalisation to qudit systems}
\label{generalisation}
The generalization of the bQKDP to qudit systems is straightforward, with the only condition of $d$ being even. It is clear from the bQKDP$_4$ and bQKDP$_6$ that a minimum of three bases are required for the secure distribution of keys. 
These bases are a must in order to detect subspace attacks as has been shown in figure (\ref{Subspace attack}).


For this reason,  the bQKDP$_d$  employs three bases. 
Similar to the illustrative protocols, we fix the basis ${\cal B}_0$ to be the computational basis. The rest of the two bases, {\it viz.}, ${\cal B}_1$ and ${\cal B}_2$ consist of states which are superpositions of two states belonging to the computational basis. The three bases are given explicitly as follows:
\begin{align}
    &{\cal B}_0\equiv \Big\{\ket{0},\ket{1},\cdots, \ket{d-1}\Big\},\nonumber\\
    &   {\cal  B}_1\equiv \Big\{\frac{1}{\sqrt{2}}\Big(\ket{0}\pm\ket{1}\Big), \frac{1}{\sqrt{2}}\Big(\ket{2}\pm\ket{3}\Big),\cdots, \frac{1}{\sqrt{2}}\Big(\ket{d-2}\pm \ket{d-1}\Big)\Big\},\nonumber\\
    &   {\cal  B}_2\equiv \Big\{\frac{1}{\sqrt{2}}\Big(\ket{1}\pm\ket{2}), \frac{1}{\sqrt{2}}\Big(\ket{3}\pm\ket{4}\Big),\cdots, \frac{1}{\sqrt{2}}\Big(\ket{d-1}\pm \ket{0}\Big)\Big\}.
\end{align}

The steps of the protocols are essentially the same as in section (\ref{QKD_four}). \\
The key generation rule, following the same idea of unambiguous state/subspace discrimination, is as follows:  
\begin{enumerate}
\item The rounds in which both Alice and Bob choose the same bases, Bob can uniquely identify the state sent by Alice. So, he obtains the full information which, in this case, is $\log_2{d}$ bits.
\item The rounds in which Alice and Bob choose different bases from either sets $\{{\cal B}_0,{\cal B}_1\}$ and $\{{\cal B}_0,{\cal B}_2\}$, a key letter is generated by employing the rule given in table (\ref{tab:rule_qudit}).


    \item  To generate key letters, Bob needs to determine uniquely the state/subspace sent by Alice. For this purpose, it is defined that only alternate states are used for key generation. The rounds in which Alice and Bob choose different bases from the set $\{{\cal B}_1,{\cal B}_2\}$, Bob - for a pair of outcomes - can determine uniquely the set from which Alice has sent states. This condition of unambiguous discrimination results in the generation of  $\frac{d}{4}$ key letters. The details of the two cases are given as follows:
\begin{itemize}
    \item For $d=4n ~(n \geq 2)$, in order to unambiguously discriminate the subspaces, Alice first discards the rounds in which she has sent the states $\frac{1}{\sqrt{2}}(\ket{2}\pm \ket{3}), \frac{1}{\sqrt{2}}(\ket{6}\pm \ket{7}), \cdots, \frac{1}{\sqrt{2}}(\ket{d-2}\pm \ket{d-1})$. This leads to unambiguous subspace discrimination, which, in turn, leads to the generation of key letters, as shown in the table (\ref{tab:rule_d} {\color{blue}(a)}) of Appendix (\ref{keyrulequdit}).
    \item For $d=4n+2~(n \geq 1)$, in order to unambiguously discriminate the subspaces, as in previous case,  Alice first discards the rounds in which alternate states are sent. In addition to this, Bob discards the rounds in which his post-measurement states are   $\frac{1}{\sqrt{2}}(\ket{d-1}\pm \ket{0})$. This leads to unambiguous subspace discrimination, which, in turn, leads to the generation of key letters, as shown in the table (\ref{tab:rule_d} {\color{blue}(b)}) of Appendix (\ref{keyrulequdit}).
\end{itemize}
\end{enumerate}

\begin{minipage}{.55\linewidth} %
\centering
\includegraphics[width=0.75\textwidth]{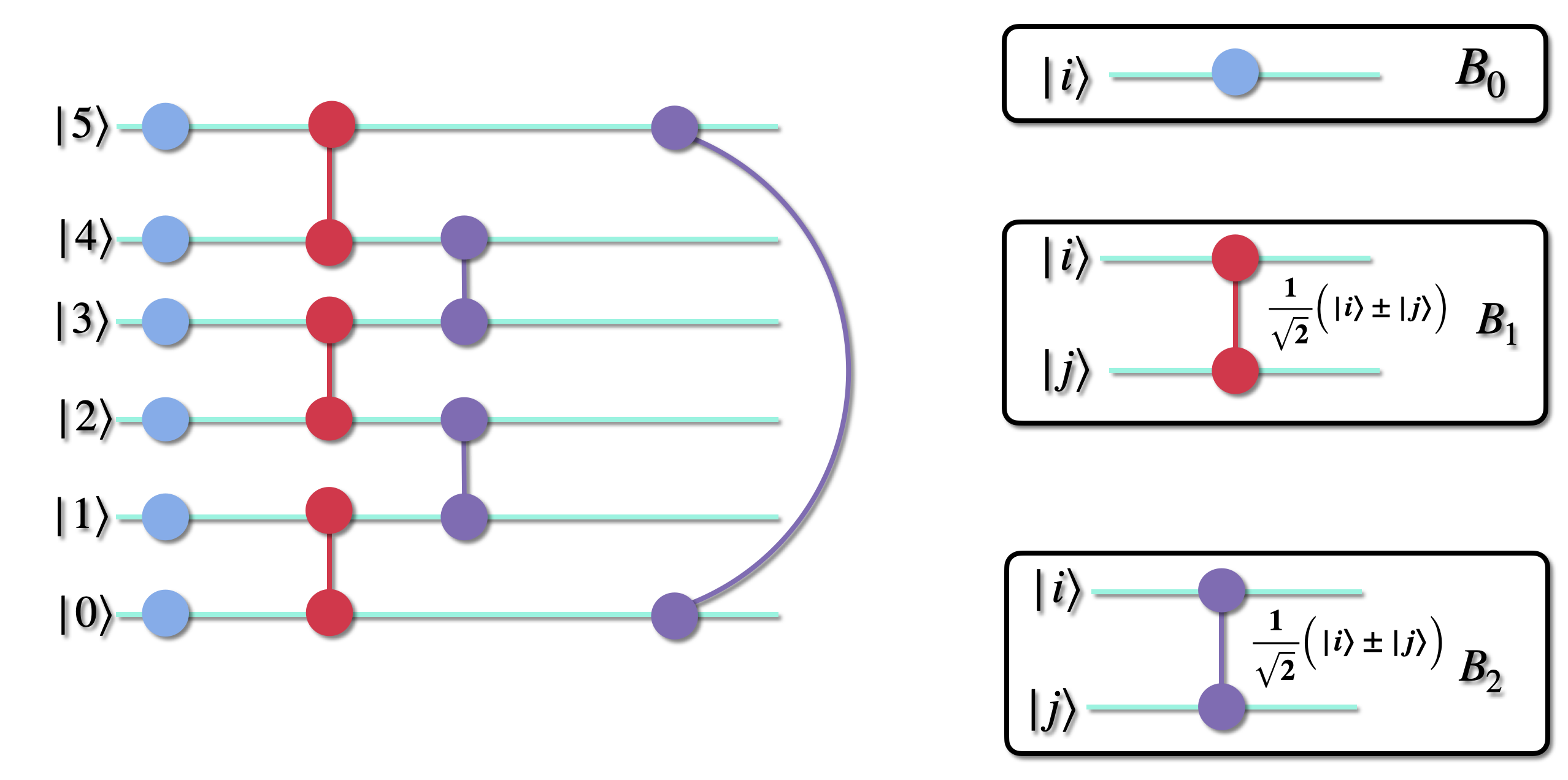}
\captionof{figure}{Schematic representation of choice of bases for the bQKDP$_6$ involving a quhex (six-dimensional) system. Since the states belonging to three bases do not allow for a reducible representation, all the subspace attacks will be detected.}\label{Subspace attack}
\end{minipage} %
\begin{minipage}{.35\linewidth} %
\centering
\resizebox{!}{1cm}{
\begin{tabular}{|c|c|c|c|c|} 
 \hline
\multirow{3}{*}{\bf Alice's basis}& \multirow{3}{2cm}{{\bf Alice's state}}&\multirow{3}{2.7cm}{{\bf Bob's basis}}& \multirow{3}{3cm}{{\bf Post-measurement state of Bob}}&\multirow{3}{*}{\bf Key symbol}\\
 &   &&& \\
 &&&&\\
     \hline\hline
\multirow{6}{*}{$\boldsymbol{{\cal B}_0}$}  & \multirow{2}{*}{ $\boldsymbol{\ket{0}/\ket{1}}$}&\multirow{6}{*}{$\boldsymbol{{\cal B}_1}$}&\multirow{2}{*}{$\boldsymbol{\frac{1}{\sqrt{2}}(\ket{0}\pm \ket{1}})$}&\multirow{2}{*}{$\boldsymbol{0}$}\\
&&&&\\
&\multirow{2}{*}{$\boldsymbol{\vdots}$}&&\multirow{2}{*}{$\boldsymbol{\vdots}$}&\multirow{2}{*}{$\boldsymbol{\vdots}$}\\
&&&&\\
&\multirow{2}{*}{$\boldsymbol{\ket{d-2}/\ket{d-1}}$}&&\multirow{2}{*}{$\boldsymbol{\frac{1}{\sqrt{2}}(\ket{d-2}\pm \ket{d-1}})$}&\multirow{2}{*}{$\boldsymbol{\dfrac{d}{2}-1}$}\\
&&&&\\
  \hline  
\multirow{6}{*}{$\boldsymbol{{\cal B}_0}$}  & \multirow{2}{*}{ $\boldsymbol{\ket{1}/\ket{2}}$}&\multirow{6}{*}{$\boldsymbol{{\cal B}_2}$}&\multirow{2}{*}{$\boldsymbol{\frac{1}{\sqrt{2}}(\ket{1}\pm \ket{2}})$}&\multirow{2}{*}{$\boldsymbol{0}$}\\
&&&&\\
&\multirow{2}{*}{$\boldsymbol{\vdots}$}&&\multirow{2}{*}{$\boldsymbol{\vdots}$}&\multirow{2}{*}{$\boldsymbol{\vdots}$}\\
&&&&\\
&\multirow{2}{*}{$\boldsymbol{\ket{d-1}/\ket{0}}$}&&\multirow{2}{*}{$\boldsymbol{\frac{1}{\sqrt{2}}(\ket{d-1}\pm \ket{0}})$}&\multirow{2}{*}{$\boldsymbol{\dfrac{d}{2}-1}$}\\
&&&&\\
  \hline  
\end{tabular}
}
\captionof{table}{Key generation rule between Alice and Bob when both measures in distinct bases.}\label{tab:rule_qudit}
\end{minipage}

\subsection{KGR}
\label{KGR_qudit}
Employing the key generation rules described above, key letters are generated in most of the rounds.  As a result, the key generation rate increases considerably as can be seen below. 

\begin{figure}[htb!]
    \centering
\includegraphics[width=0.5\textwidth]{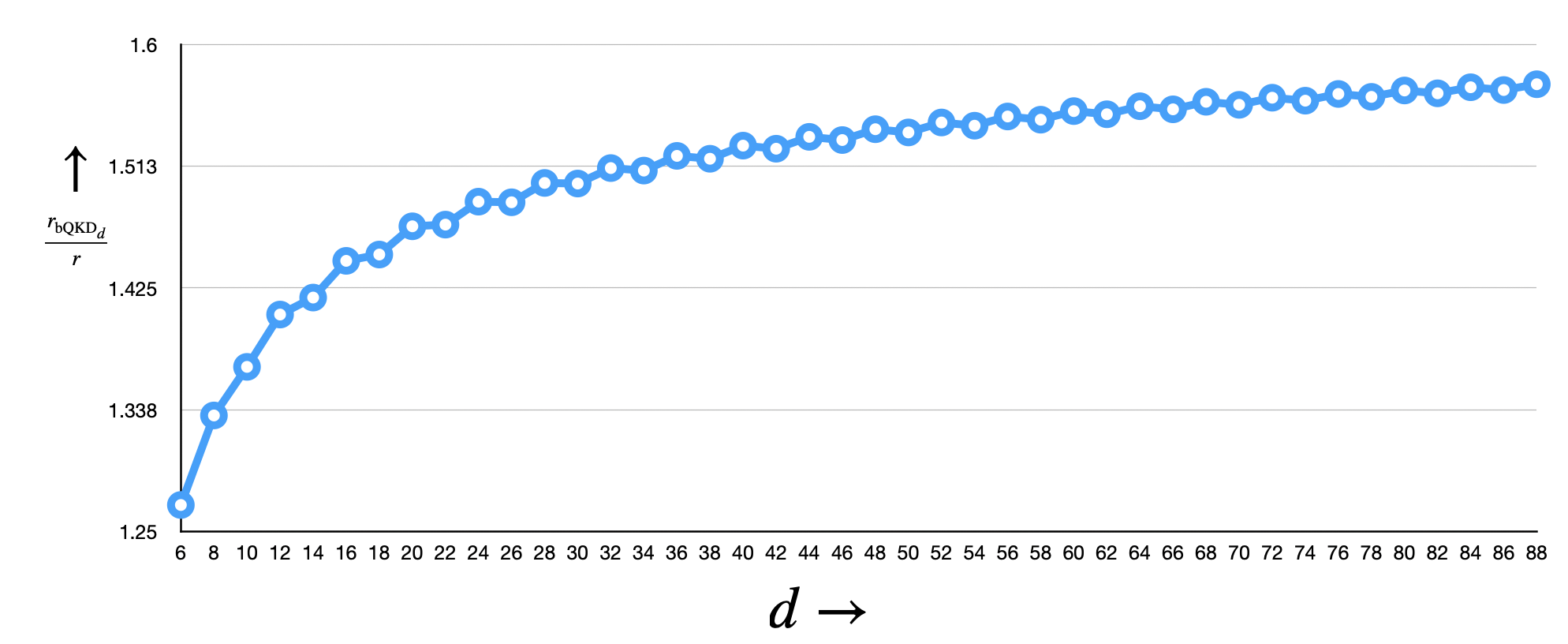}
\caption{Plot of ratio of the key generation rates of the bQKDP$_d$ (denoted  by $r_{bQKD_d}$) and that of BB84 protocol (denoted by $r$) with qudits involving only two bases ({\it viz.}, the computational basis and the Fourier basis).}
\label{Two_KGR}
\end{figure}

\noindent For $d=4n$ where $ (n\geq 2)$
\begin{align}
 r_{{\rm bQKD}_d}&=\frac{1}{3^2}\Big(3\log_2{d}+4\log_2{\frac{d}{2}}+\log_2{\frac{d}{4}}\Big){\rm ~bits~per~transmission},
\end{align}
For $d=4n+2$
\begin{align}
 r_{{\rm bQKD}_d} &=\frac{1}{3^2}\Big\{3\log_2{d}+4\log_2{\frac{d}{2}}+\frac{d-2}{d} h\Big(\frac{1}{2n},\frac{1}{2n},\frac{1}{n},\cdots,\frac{1}{n}\Big)\Big\}{\rm ~bits~per~transmission}.
\end{align}
 The symbol $h(p_1, p_2, \cdots, p_N)$ represents Shannon entropy defined as $h(p_1, p_2, \cdots, p_N)=-\sum_{i=1}^Np_i\log_2p_i$.
 The KGR of the latter is given by $r=\frac{\log_2{d}}{2}$. The value of $r_{{\rm bQKD}_d}$  is much higher than KGR of $d$-dimensional BB$84$, as is also reflected from the plot shown in figure (\ref{Two_KGR}).

The security analysis given in section (\ref{security_QKD}) of the bQKDP$_4$  can be extended to bQKDP$_d$  straightforwardly. This ensures the robustness of the protocol against various eavesdropping strategies.

\noindent  Though the protocol proposed in \cite{chau2015quantum} also employs effective qubits encoded in qudits, bQKDPs has advantages of exponentially higher KGR compared to that of \cite{chau2015quantum}.  In fact, for $d=2^n$, KGR of bQKDP is of $o(n)$, while the KGR of the protocol proposed in \cite{chau2015quantum} is $\approx\frac{1}{\binom{d}{2}^2}\sim 4^{-2n}$.

Since this idea provides an advantage in QKDPs, it becomes worthwhile to see what advantage this idea offers when employed in SQKDPs. We study it in the subsequent section. Please note that for the purpose of an uncluttered discussion, we present a brief recap of the semi-quantum key distribution protocol proposed by Boyer \cite{Boyer07} in Appendix (\ref{SQKD_Boyer}). 

\section{Boosted semi-quantum key distribution (bSQKD)}
\label{boosted SQKD}

In this section, we propose a boosted SQKDP (bSQKDP)  and demostrate it  with ququart systems.\\
\subsection{The bSQKDP$_4$}
In this protocol, Alice has all the quantum capabilities, i.e., she can prepare a state and perform measurements on it in any  basis. However, as Bob is classical, he can perform a measurement in only one of the basis states, i.e.,  the computational basis. 
 Alice has three sets of four-dimensional basis states (which is the same as given in equation (\ref{basis_ququart})).  As Bob is classical, he can measure  only in the computational basis i.e., $B_0$.
 The steps for the protocol are as follows:

\begin{enumerate}
    \item Alice randomly prepares one of the states from any of the three basis sets and sends it to Bob.
    \item Bob exercises two choices with equal probability:
    \begin{itemize}
        \item He performs the measurement in the basis $B_0$ and sends the post-measurement state to Alice.
        \item He does not measure and simply sends the state back to Alice.
    \end{itemize}
    \item Alice measures the received state in the same basis in which she has prepared the state.
    \item This completes one round. The same process is repeated for many rounds.
    \item After a sufficient number of rounds, Bob reveals the rounds in which he performs the measurement.
    \item Alice uses data from other rounds to check the presence of an eavesdropper.
    \item In the absence of Eve, Alice reveals the bases employed in each round.
    \item The rounds in which Bob performs a measurement constitute a key.
\end{enumerate}

The rounds in which Alice sends a state from the basis $B_0$, two bits of information is generated. However, if Alice sends states from basis $B_1$ or $B_2$, $1$ bit of information is generated as in bQKDP$_4$. This clearly indicates that the data of no rounds in which Bob measures is discarded in contrast to conventional SQKDPs.

\begin{figure}
     \centering     \includegraphics[width=9cm]{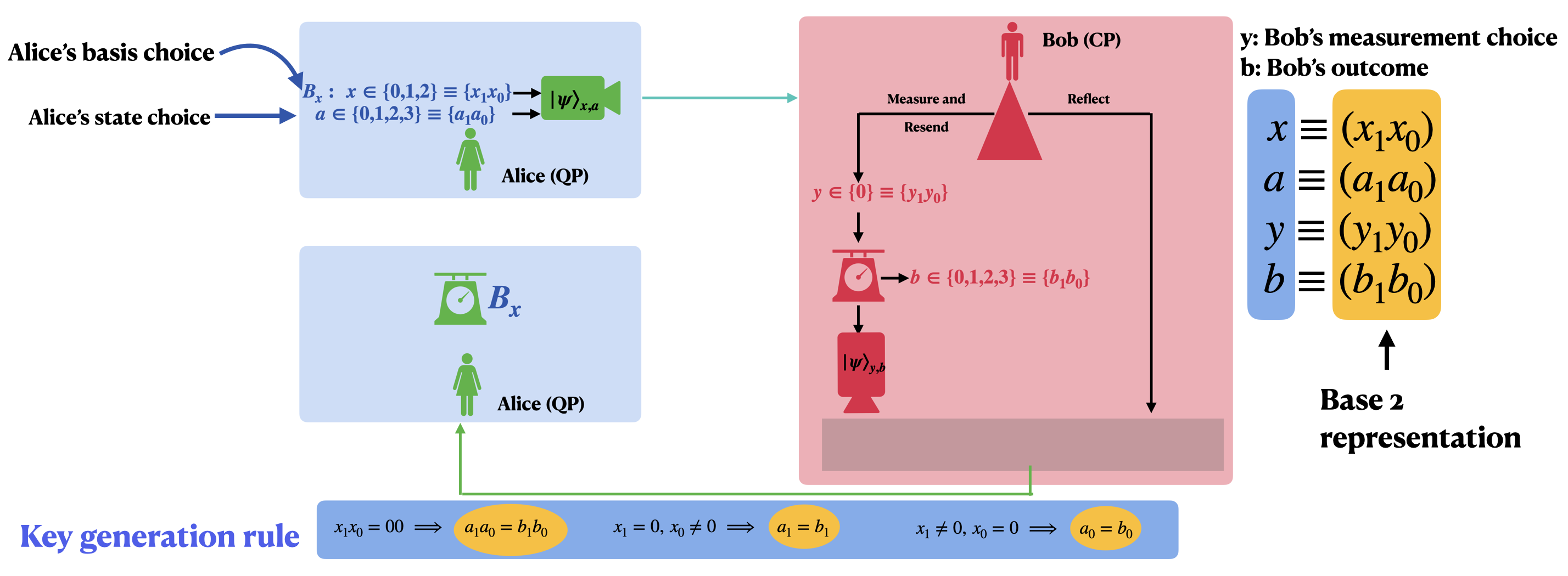}
     \caption{Pictorial representation of bSQKDP$_4$.}
     \label{ESQKD_sooryansh}
 \end{figure}

 The protocol has been schematically shown in figure (\ref{ESQKD_sooryansh}) reflecting how different bases leads to key generation. 
\subsection{KGR}
From the above discussion, it is clear that a bit for the key is generated whenever Bob performs a measurement in the basis $B_0$, which happens with a probability of $1/2$. The probability for Alice to choose any one of the bases set is also equal and is $1/3$. Therefore, the KGR ($r_{{\rm bSQKD}_4}$) for this protocol is: 
\begin{align}
    r_{{\rm bSQKD}_4}&= \frac{1}{3}\times\frac{1}{2}\times 2 +\frac{1}{3}\times\frac{1}{2}\times 1+\frac{1}{3}\times\frac{1}{2}\times 1=\frac{2}{3}=0.66 ~\rm{bits}.
\end{align}
In the protocol, we have assumed that probability of Bob's two operations, {\it viz.}, reflect and measure, is the same. However, this can be tweaked such that Bob measure with probability $q>\frac{1}{2}$ which increases the KGR to  $r_{{\rm bSQKD}_4}=\frac{4}{3}q$.

We now move on to present the robustness of the protocol against various eavesdropping strategies.


\subsection{Robustness of the protocol}
\label{Robustness_SQKD}
In this section, we discuss the robustness of the bSQKDP$_4$.  Like other SQKDPs, bSQKDP$_4$ is also a two-way communication protocol. Exploiting this, Eve may eavesdrop either in both ways or only in one way. The security against a one-way eavesdropping attack follows directly from the security of the bQKDP. So, in what follows, we discuss the robustness of the protocol against two-way entangling attacks.  
\subsubsection{Two-way entangling attack}
In such an attack, Eve may try to interact with traveling states with her ancillae. Let $U_F$ and $U_B$ be the two transformations that Eve employs to interact with states traveling from Alice to Bob and Bob to Alice respectively. The actions of these transformations can be expressed as:
 \begin{align}\label{eq:unitary}
     U_F\ket{i}_T\ket{0}\to \sum_{j=0}^3 \ket{j}_T\ket{E_{ij}},~ U_B\ket{j}_T\ket{0}\to \sum_{k=0}^3\ket{k}_T\ket{F_{jk}}
 \end{align}
 In the protocol, since Bob is classical, he either measures in the basis $B_0$ or he simply sends the received states to Alice. We discuss the two cases one-by-one.
 \subsubsection*{A. Bob performs a measurement}
 In the protocol, whenever Bob performs measurement, key letters are generated. Alice and Bob, by comparing a subset of this data, can detect Eve's tampering in both the directions. That is to say, Eve's tampering when states travel from Alice to Bob and Bob to Alice can be detected.\\

 \noindent{\bf (a) Alice sends states from the basis $B_0$:} Whenever Alice sends a state say $\ket{i}$ from  the basis $B_0$, Bob gets the correct state only with the probability, $\norm{\ket{E_{ii}}}^2$. Given Bob gets the state sent by Alice, Alice gets the same state with a probability $\norm{\ket{F_{ii}}}^2$. So, both Alice and Bob get the same state as initially sent by Alice with a probability $p=\norm{\ket{E_{ii}}}^2\norm{\ket{F_{ii}}}^2$. Thus, Eve's interventions are detected  with a probability $(1-p)$ when Alice and Bob compare a subset of their outcomes.

 \noindent{\bf (b) Alice sends states from the basis $B_1$:} The effect of Eve's interaction on the states of the basis $B_1$ traveling from Alice to Bob can be described as follows: 
  \begin{align}
    &  \frac{1}{\sqrt{2}}\big(\ket{m}\pm\ket{n}\big)\ket{0}\xrightarrow[]{U_F}\frac{1}{\sqrt{2}}\sum_{j=0}^3\ket{j}\Big(\ket{E_{mj}}\pm \ket{E_{nj}}\Big),
\end{align}
where $m=0,~ n=1{\rm~ or~} m=2,~n=3$.
So, Bob gets the correct result if his post-measurement state is either $\ket{m}$ or $\ket{n}$. He gets it with probability $\frac{1}{2}\big(\norm{\ket{E_{mm}}\pm\ket{E_{nm}}}^2+\norm{\ket{E_{mn}}\pm\ket{E_{nn}}}^2\big)$.
Since Alice measure in the same basis in which she has prepared the state, following equation (\ref{eq:unitary}), she gets the correct outcomes with probability $\frac{1}{2}\norm{\ket{F_{mm}}+\ket{F_{mn}}}^2+\frac{1}{2}\norm{\ket{F_{mm}}-\ket{F_{mn}}}^2+\frac{1}{2}\norm{\ket{F_{nn}}+\ket{F_{nm}}}^2+\frac{1}{2}\norm{\ket{F_{nn}}-\ket{F_{nm}}}^2$.

\begin{table}[!htb]
        \begin{minipage}{.65\linewidth}
\resizebox{!}{1.23cm} {
    \centering    
\begin{tabular}{ |c|c| c|c| c|c|}
\hline
\multirow{4}{2cm}{\bf State sent by Alice}&\multirow{4}{3cm}{\centering\bf  Bob's post-measurement state corresponding to correct outcome}&\multirow{4}{3.5cm}{\centering \bf Probability of correct outcome at Bob's end} &\multirow{4}{3.5cm}{\centering\bf Probability of correct outcome at Alice's end}&\multirow{4}{3cm}{\bf Probability of getting correct outcome}&\multirow{4}{1.85cm}{\bf Probability of detecting Eve}\\
 &&&&& \\
 &&&&&\\
&&&&&\\\hline
\multirow{4}{*}{{\large${\bf \ket{i}}$}} &\multirow{4}{*}{\large${\bf \ket{i}}$}&\multirow{4}{*}{\large${\bf \norm{\ket{E_{ii}}}^2}$}&\multirow{4}{*}{\large${\bf \norm{\ket{F_{ii}}}^2}$}&\multirow{4}{*}{\large${\bf \norm{\ket{E_{ii}}}^2\norm{\ket{F_{ii}}}^2}$}&\multirow{4}{*}{\large${\bf 1-\norm{\ket{E_{ii}}}^2\norm{\ket{F_{ii}}}^2}$}\\
&&&&&\\
&&&&&\\
&&&&&\\
\hline
\multirow{8}{*}{\large${\bf \ket{i}+\ket{j}}$} &\multirow{4}{*}{\large${\bf \ket{i}}$}&\multirow{4}{*}{\large${\bf p_{+i}=\frac{1}{2}\norm{\ket{E_{ii}}+\ket{E_{ji}}}^2}$}&\multirow{4}{*}{\large${\bf p_{i}=\frac{1}{2}\norm{\ket{F_{ii}}+\ket{F_{ij}}}^2+\frac{1}{2}\norm{\ket{F_{ii}}-\ket{F_{ij}}}^2}$}&\multirow{4}{*}{{\large ${\bf p_{+i}p_{i}}$}}&\multirow{8}{*}{\large${\bf  1-p_{+i}p_{i}-p_{+j}p_{j}}$}\\
&&&&&\\
&&&&&\\
&&&&&\\
&\multirow{4}{*}{\large${\bf \ket{j}}$}&\multirow{4}{*}{\large${\bf p_{+j}=\frac{1}{2}\norm{\ket{E_{ij}}+\ket{E_{jj}}}^2}$}&\multirow{4}{*}{\large${\bf p_{j}=\frac{1}{2}\norm{\ket{F_{jj}}+\ket{F_{ji}}}^2+\frac{1}{2}\norm{\ket{F_{jj}}-\ket{F_{ji}}}^2}$}&\multirow{4}{*}{\large${\bf p_{+j}p_{j}}$}&\\
&&&&&\\
&&&&&\\
&&&&&\\\hline
\multirow{8}{*}{\large${\bf \ket{i}-\ket{j}}$} &\multirow{4}{*}{\large${\bf \ket{i}}$}&\multirow{4}{*}{\large${\bf p_{-i}=\frac{1}{2}\norm{\ket{E_{ii}}-\ket{E_{ji}}}^2}$}&\multirow{4}{*}{\large${\bf p_{i}=\frac{1}{2}\norm{\ket{F_{ii}}-\ket{F_{ij}}}^2+\frac{1}{2}\norm{\ket{F_{ii}}+\ket{F_{ij}}}^2}$}&\multirow{4}{*}{\large${\bf p_{-i}p_{i}}$}&\multirow{8}{*}{\large${\bf  1-p_{-i}p_{i}-p_{-j}p_{j}}$}\\
&&&&&\\
&&&&&\\&&&&&\\
&\multirow{4}{*}{\large${\bf \ket{j}}$}&\multirow{4}{*}{\large${\bf p_{-j}=\frac{1}{2}\norm{\ket{E_{ij}}-\ket{E_{jj}}}^2}$}&\multirow{4}{*}{\large${\bf p_{j}=\frac{1}{2}\norm{\ket{F_{jj}}-\ket{F_{ji}}}^2+\frac{1}{2}\norm{\ket{F_{jj}}+\ket{F_{ji}}}^2}$}&\multirow{4}{*}{\large${\bf p_{-j}p_{j}}$}&\\
&&&&&\\
&&&&&\\
&&&&&\\\hline
  \end{tabular}
    }
   \caption*{(a) }
    \end{minipage}%
    \begin{minipage}{.35\linewidth}
 \resizebox{!}{0.78cm}{
    \centering   
\begin{tabular}{ |c| c|c|}
\hline
\multirow{3}{3cm}{{\bf State sent by Alice}}&\multirow{3}{3.5cm}{{\bf Probability with which Alice gets correct outcome }}&\multirow{3}{1.85cm}{{\bf Probability of detecting Eve}}\\
 && \\
&&\\\hline
\multirow{4}{*}{{\large${\bf \ket{i}}$}} &\multirow{4}{*}{\large${\bf p_i=\norm{\sum_{j=0}^3\ket{E_{ij}}\ket{F_{ji}}}^2}$}&\multirow{4}{*}{\large${\bf 1-p_i}$}\\
&&\\
&&\\
&&\\
\hline
\multirow{4}{*}{\large${\bf \ket{i}+\ket{j}}$} &\multirow{4}{*}{{\large ${\bf p_+=\frac{1}{2}\norm{\sum_{k=0}^3\sum_{m,n=i,j}\ket{E_{mk}}\ket{F_{kn}}}^2}$}}&\multirow{4}{*}{\large${\bf  1-p_+}$}\\
&&\\
&&\\
&&\\\hline
\multirow{4}{*}{\large${\bf \ket{i}-\ket{j}}$} &\multirow{4}{*}{\large${\bf p_-=\frac{1}{2}\norm{\sum_{k=0}^3\sum_{m,n=i,j}(-1)^{m+n}\ket{E_{mk}}\ket{F_{kn}}}^2}$}&\multirow{4}{*}{\large${\bf  1-p_-}$}\\
&&\\
&&\\
&&\\\hline
  \end{tabular}
    }
    \caption*{(b)}
    \end{minipage} 
      \caption{Probability of  detection of eavesdropping when (a) Bob performs measurement, and (b) when Bob performs Reflect operation.}
        \label{tab:Comparison_QKD}
\end{table}

The same analysis holds when Alice sends states from the basis $B_2$. In this case, $m=0,~n=3$ or $m=1,~n=2$.
 The detailing of the same is given in table (\ref{tab:Comparison_QKD} {\color{blue} (a)}).

\subsubsection*{B. Bob performs reflect operation}

In the absence of an eavesdropper, the rounds in which Bob performs the reflect  operation, Alice would get the same post-measurement state as she had sent. However, due to Eve's interventions, this will not be the case, reflecting her presence and thus making these protocols robust against eavesdropping attacks.

Since Bob performs reflect operation, the action of Eve's attack in such case can be expressed as:
\begin{align}
    &U_BU_F\ket{i}\ket{0}_E\ket{0}_B\rightarrow U_B\sum_{j=0}^3\ket{j}\ket{E_{ij}}\ket{0}_B\rightarrow \sum_{j,k=0}^3 \ket{k}\ket{E_{ij}}\ket{F_{jk}}
\end{align}
As is clear from the above equation, whenever Alice sends a state from the basis $B_0$, she  gets the correct result with probability $p=\norm{\sum_{j=0}^3\ket{E_{ij}}\ket{F_{ji}}}^2$. Thus, Eve's interventions get detected with a probability $(1-p)$.

\noindent{\textbf{Alice sends state from the basis $B_1$:}}
Suppose that Alice sends the state $\frac{1}{\sqrt{2}}(\ket{0}+\ket{1})$. Since Bob performs a reflect operation, the effect of Eve's interactions on the state can be expressed as: 
\begin{align}
        U_BU_F\frac{1}{\sqrt{2}}\Big(\ket{0}+\ket{1}\Big)\ket{0}\ket{0}
    &=\frac{1}{\sqrt{2}}\Big(\sum_{i,j=0}^3\ket{j}\big(\ket{E_{0i}}+\ket{E_{1i}}\big)\ket{F_{ij}}\Big)
\end{align}
Quite clearly, this state is nonorthogonal to the states $\frac{1}{\sqrt{2}}(\ket{0}-\ket{1})$ and $\frac{1}{\sqrt{2}}(\ket{2}\pm\ket{3})$, which implies that Alice's outcome will not always correspond to the initial state that she had sent. This signals eavesdropping. 

The same analysis holds when Alice sends state from the basis $B_2$. The probability of Eve's detection is given in table ((\ref{tab:Comparison_QKD} {\color{blue} (b)})). This concludes our discussion on the robustness of bSQKDP$_4$ against eavesdropping.

The generalization of bSQKDPs to effective qubits encoded in high-dimensional systems is straightforward. The steps of the protocol essentially remains the same as of bSQKDP$_4$. The same bases as in bQKDP$_d$ (given in section (\ref{generalisation})) can be employed. 
We now briefly discuss the experimental setup for the implementation of bQKDP$_4$ and bSQKDP$_4$ protocol. \\

 \noindent {\it Experimental implementation:}\\
 
 \noindent The bQKDP$_4$ and bSQKDP$_4$  (proposed in sections (\ref{QKD_four}) and (\ref{boosted SQKD})) may be implemented with OAM modes of light. For this, the states belonging to the computational basis may be identified with  the Laguerre-Gauss modes of light. These states can be generated by passing Gaussian modes of light through appropriate phase masks.  The states belonging to other two basis involve coherent superpositions of two Laguerre Gauss modes respectively which can be realized with spatial light modulators \cite{PhysRevApplied.11.064058}. In fact, effective qubits encoded in qudits have been generated with fidelities of $\sim 99\%$ \cite{ding2017high}, thereby making QKDPs involving them  experimentally more feasible. One can also employ a digital micro-mirror device as has been done in \cite{mirhosseini2015high} for preparing the two bases. In fact, since the bQKDP and bSQKDP do not employ entangled states, they may be implemented with faint coherent pulses. To prepare faint coherent pulses, an attenuator can be employed. In fact, two attenuators with different degrees of attenuation  can be employed to make the protocols resilient against photon-number-splitting (PNS) attacks (discussed in detail in section (\ref{PNS})). Measurements of Laguerre Gauss modes (and superposition thereof) can be realized by employing a spatial light modulator (SLM) and detectors or with the techniques  presented in \cite{mirhosseini2013efficient}.\\

\section{Conclusion}
\label{conclusion} 
 In this paper, we have proposed bQKDPs  and bSQKDPs by employing effective qubits encoded in higher dimensional systems and have shown their robustness against various eavesdropping startegies. The protocols have several advantages: (i) higher KGR, (ii) ease in preparation of states, (iii) realizable with linear optics only, (iv) easy generalizability to other protocols, {\it viz.}, quantum dialogue and quantum key agreement, etc. Another direction in which our protocols can be generalized is by increasing the number of participants (which is chosen to be two in this paper, for the sake of simplicity). In fact, employing this strategy in a quantum network would constitute an interesting study. Furthermore, the protocols can also be generalized to those quantum networks, in which different parties are divided into different subsets constituting different {\it layers} \cite{pivoluska2018layered}. The protocols can be made robust against side-channel attacks by extending them to the measurement-device-independent setting. A more rigorous information-theoretic security analysis of the proposed protocols, taking into account the finite key analysis \cite{tomamichel2017largely}\footnote{We thank the referee for bringing this reference to our attention.},  deserve serious attention, which we hope to take up in the near future.



\begin{appendices}
\end{appendices}

\section*{Acknowledgements}
{\color{blue} We thank the referees for important suggestions, which have helped us improve the quality of the manuscript.} Rajni and Sooryansh thank UGC and CSIR (File no.: 09/086(1278)/ 2017-EMR-I) for funding their research in the initial stages of the work.
\section*{Data availability statement}
Data sharing is not applicable to this article as no datasets were generated or analyzed during the current study.
\section*{Disclosures}
The authors declare no conflicts of interest.
\appendix

\section{bQKDP with quhex systems (bQKD$_6$)}
\label{QKD_six}
In this section, we present a bQKDP with a quhex system. Let there be two participants, {\it viz.}, Alice and Bob who wish to share a secret key. 

\noindent{\textbf{\textit{Aim:}}} distribution of a key with minimal discarding of data.

\noindent{\textbf{\textit{Resources:}}} effective qubits in quhex systems with three bases $B_0, B_1, B_2$
\begin{align}
    &B_0\equiv\Big\{\ket{0},\ket{1},\ket{2},\ket{3},\ket{4},\ket{5}\Big\};\nonumber\\
    &B_1\equiv\Big\{\frac{1}{\sqrt{2}}\Big(\ket{0}\pm\ket{1}\Big),\frac{1}{\sqrt{2}}\Big(\ket{2}\pm\ket{3}\Big),\frac{1}{\sqrt{2}}\Big(\ket{4}\pm\ket{5}\Big)\Big\};\nonumber\\
      &B_2\equiv\Big\{\frac{1}{\sqrt{2}}\Big(\ket{1}\pm\ket{2}\Big),\frac{1}{\sqrt{2}}\Big(\ket{3}\pm\ket{4}\Big),\frac{1}{\sqrt{2}}\Big(\ket{5}\pm\ket{0}\Big)\Big\}.
\label{bases_six_sooryansh}
\end{align}
The steps of the protocol are the same as those of bQKDP$_4$, given in section (\ref{QKD_four}). The subtle difference is that data corresponding to none of bases choices are discarded in contrast to bQKDP$_4$. \\

\noindent{\textbf{Key generation rule:}}
\begin{itemize}
    \item The rounds in which Alice and Bob measure in the same bases,  $6$ key letters are generated with equal probabilities, sharing $\log_2{6}$ bits of information.
    \item The rounds in which either Alice or Bob chooses the basis $B_0$, $3$ key symbols are generated with equal probability as given in table (\ref{tab:bqkd6} {\color{blue}(a)}).
    \item The third case is unique to this protocol. This corresponds to  the rounds in which Alice and Bob choose the bases $B_1$ and $B_2$ respectively or vice-versa. In contrast to bQKDP$_4$, in bQKDP$_6$, these rounds contribute to key generation. The key generation rule for the same is given in table (\ref{tab:bqkd6} {\color{blue}(b)}).
\end{itemize}
\begin{table}[!htb]
    \label{tab:distinct_1}
    \begin{minipage}{.5\linewidth}
      \centering
        \resizebox{!}{1.1cm}{
\begin{tabular}{|c|c|c|c|c|} 
 \hline
\multirow{3}{*}{\bf Alice's basis}& \multirow{3}{2cm}{{\bf Alice's state}}&\multirow{3}{2.7cm}{{\bf Bob's basis}}& \multirow{3}{3cm}{{\bf Post-measurement state of Bob}}&\multirow{3}{*}{\bf Key symbol}\\
 &   &&& \\
 &&&&\\
     \hline\hline
\multirow{6}{*}{${\bf {\cal B}_0}$}  & \multirow{2}{*}{ ${\bf \ket{0}/\ket{1}}$}&\multirow{6}{*}{${\bf {\cal B}_1}$}&\multirow{2}{*}{${\bf \frac{1}{\sqrt{2}}\big(\ket{0}\pm \ket{1}\big)}$}&\multirow{2}{*}{${\bf 0}$}\\
&&&&\\
&\multirow{2}{*}{${\bf \ket{2}/\ket{3}}$}&&\multirow{2}{*}{${\bf \frac{1}{\sqrt{2}}\big(\ket{2}\pm \ket{3}\big)}$}&\multirow{2}{*}{${\bf 1}$}\\
&&&&\\
&\multirow{2}{*}{${\bf \ket{4}/\ket{5}}$}&&\multirow{2}{*}{${\bf \frac{1}{\sqrt{2}}\big(\ket{4}\pm \ket{5}\big)}$}&\multirow{2}{*}{${\bf 2}$}\\
&&&&\\
  \hline
\multirow{6}{*}{${\bf {\cal B}_0}$}  & \multirow{2}{*}{ ${\bf \ket{1}/\ket{2}}$}&\multirow{6}{*}{${\bf {\cal B}_2}$}&\multirow{2}{*}{${\bf \frac{1}{\sqrt{2}}\big(\ket{1}\pm \ket{2}\big)}$}&\multirow{2}{*}{${\bf 0}$}\\
&&&&\\
&\multirow{2}{*}{${\bf \ket{3}/\ket{4}}$}&&\multirow{2}{*}{${\bf \frac{1}{\sqrt{2}}\big(\ket{3}\pm \ket{4}\big)}$}&\multirow{2}{*}{${\bf 1}$}\\
&&&&\\
&\multirow{2}{*}{${\bf \ket{5}/\ket{0}}$}&&\multirow{2}{*}{${\bf \frac{1}{\sqrt{2}}\big(\ket{5}\pm \ket{0}\big)}$}&\multirow{2}{*}{${\bf 2}$}\\
&&&&\\
  \hline
\end{tabular}
}
      \caption*{(a)}
    \end{minipage}%
    \begin{minipage}{.5\linewidth}
      \centering
     \resizebox{!}{1.1cm}{
\begin{tabular}{|c|c|c|c|c|} 
 \hline
 \multirow{3}{3cm}{{ \bf Alice's  basis }}&\multirow{3}{3cm}{{\bf Alice's state}}&\multirow{3}{3cm}{{\bf Bob's  basis }}& \multirow{3}{3cm}{{\bf Post-measurement state of Bob (in the basis $B_2$)}}&\multirow{3}{*}{\bf Key letter}\\
   &&&& \\
 &&&&\\
     \hline\hline
\multirow{6}{*}{\large ${\bf B_1}$}& \multirow{6}{*}{ \large${\bf \frac{1}{\sqrt{2}}\big(\ket{0}\pm\ket{1}\big)}$}&\multirow{6}{*}{\large ${\bf B_2}$}&\multirow{3}{*}{\large${\bf \frac{1}{\sqrt{2}}\big(\ket{1}\pm \ket{2}\big)}$}&\multirow{3}{*}{\large${\bf 0}$}\\
&&&&\\
&&&&\\
&&&\multirow{3}{*}{\large${\bf \frac{1}{\sqrt{2}}\big(\ket{5}\pm \ket{0}\big)}$}&\multirow{3}{*}{\large${\bf \perp}$}\\
&&&&\\
&&&&\\
  \hline
\multirow{6}{*}{\large ${\bf B_1}$}&\multirow{6}{*}{\large ${\bf \frac{1}{\sqrt{2}}\big(\ket{2}\pm\ket{3}\big)}$}&\multirow{6}{*}{\large ${\bf B_2}$}&\multirow{3}{*}{\large${\bf \frac{1}{\sqrt{2}}\big(\ket{1}\pm \ket{2}\big)}$}&\multirow{6}{*}{\large${\bf \perp}$}\\
&&&&\\
&&&&\\
&&&\multirow{3}{*}{\large${\bf \frac{1}{\sqrt{2}}\big(\ket{3}\pm \ket{4}\big)}$}&\\
&&&&\\
&&&&\\\hline
\multirow{6}{*}{\large ${\bf B_1}$}&\multirow{6}{*}{\large${\bf \frac{1}{\sqrt{2}}\big(\ket{4}\pm\ket{5}\big)}$}&\multirow{6}{*}{\large ${\bf B_2}$}&\multirow{3}{*}{\large${\bf \frac{1}{\sqrt{2}}\big(\ket{3}\pm \ket{4}\big)}$}&\multirow{3}{*}{\large${\bf 1}$}\\
&&&&\\
&&&&\\
&&&\multirow{3}{*}{\large${\bf \frac{1}{\sqrt{2}}\big(\ket{5}\pm \ket{0}\big)}$}&\multirow{3}{*}{\large${\bf \perp}$}\\
&&&&\\
&&&&\\
  \hline
\end{tabular}
}
          \caption*{(b)}     
    \end{minipage} 
        \caption{ Key generation rule when Alice and Bob choose distinct basis (a)   from the sets $\{B_0, B_1\}$ or $\{B_0, B_2\}$, and (b) from the set $\{B_1, B_2\}$. The symbol $\perp$ corresponds to the rounds which are discarded. }
        \label{tab:bqkd6}
\end{table}
\subsection{KGR}   
Alice and Bob have three bases which they choose with equal probability. From the key generation rule discussed above, for each basis choice, key letters are generated with equal probability. Therefore, the KGR of the protocol is,
\begin{equation}
    r_{{\rm bQKD}_6}=\frac{1}{3^2}\times\big(\log_2{6}\times 3+\log_2{3}\times 4+\frac{2}{3}\times \log_22\big)\approx  1.64 {\rm ~bits~per~transmission.}
\end{equation}
At this juncture, we wish to point out that there is an increase in KGR as compared to the BB84 protocol with a quhex system. The KGR of the latter is $\frac{\rm log_2 6}{2}\approx 1.29$ bit per transmission.

\section{Key generation rule for the rounds in which Alice chooses basis ${\cal B}_1$ and Bob measures in basis ${\cal B}_2$}
\label{keyrulequdit}
In this appendix, we present tables (\ref{tab:rule_d} {\color{blue} (a)}) and (\ref{tab:rule_d} {\color{blue} (b)}). In these tables, we compactly show the key generation rule when Alice chooses a state from the basis $ {\cal B}_1$ and Bob makes a measurement in the basis $ {\cal B}_2$ in bQKDP$_d$ (presented in section (\ref{generalisation})). 
\begin{table}[!htb]
    
    \begin{minipage}{.5\linewidth}
      \centering
       \resizebox{!}{3cm}{
\begin{tabular}{|c|c|c|} 
 \hline
 \multirow{3}{3cm}{{\bf Alice's state from the basis ${\cal B}_1$}}& \multirow{3}{3cm}{{\bf Post-measurement state of Bob (in the basis ${\cal B}_2$)}}&\multirow{3}{*}{\bf Key letter}\\
   && \\
 &&\\
     \hline\hline
 \multirow{4}{*}{ ${\bf \frac{1}{\sqrt{2}}(\ket{0}\pm\ket{1})}$}&\multirow{2}{*}{${\bf \frac{1}{\sqrt{2}}(\ket{d-1}\pm \ket{0})}$}&\multirow{4}{*}{${\bf 0}$}\\
&&\\
&\multirow{2}{*}{${\bf \frac{1}{\sqrt{2}}(\ket{1}\pm \ket{2})}$}&\\
&&\\
  \hline
\multirow{4}{*}{ ${\bf \frac{1}{\sqrt{2}}(\ket{2}\pm\ket{3})}$}&\multirow{2}{*}{${\bf \frac{1}{\sqrt{2}}(\ket{1}\pm \ket{2})}$}&\multirow{4}{*}{${\bf \perp}$}\\
&&\\
&\multirow{2}{*}{${\bf \frac{1}{\sqrt{2}}(\ket{3}\pm \ket{4})}$}&\\
&&\\\hline
\multirow{4}{*}{${\bf \frac{1}{\sqrt{2}}(\ket{4}\pm\ket{5})}$}&\multirow{2}{*}{${\bf \frac{1}{\sqrt{2}}(\ket{3}\pm \ket{4})}$}&\multirow{4}{*}{${\bf 1}$}\\
&&\\
&\multirow{2}{*}{${\bf \frac{1}{\sqrt{2}}(\ket{5}\pm \ket{6})}$}&\\
&&\\
  \hline
  \multirow{4}{*}{${\bf \frac{1}{\sqrt{2}}(\ket{6}\pm\ket{7})}$}&\multirow{2}{*}{${\bf \frac{1}{\sqrt{2}}(\ket{5}\pm \ket{6})}$}&\multirow{4}{*}{${\bf \perp}$}\\
&&\\
&\multirow{2}{*}{${\bf \frac{1}{\sqrt{2}}(\ket{7}\pm \ket{8})}$}&\\
&&\\
  \hline
   \multirow{4}{*}{${\bf\vdots}$}&\multirow{4}{*}{${\bf\vdots}$}&\multirow{4}{*}{${\bf\vdots}$}\\
&&\\
&&\\
&&\\
  \hline
     \multirow{4}{*}{${\bf \frac{1}{\sqrt{2}}(\ket{d-4}\pm\ket{d-3})}$}&\multirow{2}{*}{${\bf \frac{1}{\sqrt{2}}(\ket{d-5}\pm \ket{d-4})}$}&\multirow{4}{*}{${\bf \frac{d}{4}-1}$}\\
&&\\
&\multirow{2}{*}{${\bf \frac{1}{\sqrt{2}}(\ket{d-3}\pm \ket{d-2})}$}&\\
&&\\
  \hline
   \multirow{4}{*}{${\bf \frac{1}{\sqrt{2}}(\ket{d-2}\pm\ket{d-1})}$}&\multirow{2}{*}{${\bf \frac{1}{\sqrt{2}}(\ket{d-3}\pm \ket{d-2})}$}&\multirow{4}{*}{${\bf \perp}$}\\
&&\\
&\multirow{2}{*}{${\bf \frac{1}{\sqrt{2}}(\ket{d-1}\pm \ket{0})}$}&\\
&&\\
  \hline
\end{tabular}
}
\caption*{(a)}
    \end{minipage}%
    \begin{minipage}{.5\linewidth}
      \centering
 \resizebox{!}{3cm}{
\begin{tabular}{|c|c|c|} 
 \hline
 \multirow{3}{3cm}{{\bf Alice's state from the basis ${\cal B}_1$}}& \multirow{3}{3cm}{{\bf Post-measurement state of Bob (in the basis ${\cal B}_2$)}}&\multirow{3}{*}{\bf Key letter}\\
   && \\
 &&\\
     \hline\hline
 \multirow{4}{*}{ ${\bf \frac{1}{\sqrt{2}}\Big(\ket{0}\pm\ket{1}\Big)}$}&\multirow{2}{*}{${\bf \frac{1}{\sqrt{2}}\Big(\ket{d-1}\pm \ket{0}\Big)}$}&\multirow{2}{*}{${\bf \perp}$}\\
&&\\
&\multirow{2}{*}{${\bf \frac{1}{\sqrt{2}}\Big(\ket{1}\pm \ket{2}\Big)}$}&\multirow{2}{*}{${\bf 0}$}\\
&&\\
  \hline
\multirow{4}{*}{ ${\bf \frac{1}{\sqrt{2}}\Big(\ket{2}\pm\ket{3}\Big)}$}&\multirow{2}{*}{${\bf \frac{1}{\sqrt{2}}\Big(\ket{1}\pm \ket{2}\Big)}$}&\multirow{4}{*}{${\bf \perp}$}\\
&&\\
&\multirow{2}{*}{${\bf \frac{1}{\sqrt{2}}\Big(\ket{3}\pm \ket{4}\Big)}$}&\\
&&\\\hline
\multirow{4}{*}{${\bf \frac{1}{\sqrt{2}}\Big(\ket{4}\pm\ket{5}}$}&\multirow{2}{*}{${\bf \frac{1}{\sqrt{2}}\Big(\ket{3}\pm \ket{4}\Big)}$}&\multirow{4}{*}{${\bf 1}$}\\
&&\\
&\multirow{2}{*}{${\bf \frac{1}{\sqrt{2}}\Big(\ket{5}\pm \ket{6}\Big)}$}&\\
&&\\
  \hline
  \multirow{4}{*}{${\bf \frac{1}{\sqrt{2}}\Big(\ket{6}\pm\ket{7}\Big)}$}&\multirow{2}{*}{${\bf \frac{1}{\sqrt{2}}\Big(\ket{5}\pm \ket{6}\Big)}$}&\multirow{4}{*}{${\bf \perp}$}\\
&&\\
&\multirow{2}{*}{${\bf \frac{1}{\sqrt{2}}\Big(\ket{7}\pm \ket{8}\Big)}$}&\\
&&\\
  \hline
   \multirow{4}{*}{${\bf \vdots}$}&\multirow{4}{*}{${\bf \vdots}$}&\multirow{4}{*}{${\bf \vdots}$}\\
&&\\
&&\\
&&\\
  \hline
     \multirow{4}{*}{${\bf \frac{1}{\sqrt{2}}\Big(\ket{d-4}\pm\ket{d-3}\Big)}$}&\multirow{2}{*}{${\bf \frac{1}{\sqrt{2}}\Big(\ket{d-5}\pm \ket{d-4}\Big)}$}&\multirow{4}{*}{${\bf \perp}$}\\
&&\\
&\multirow{2}{*}{${\bf \frac{1}{\sqrt{2}}\Big(\ket{d-3}\pm \ket{d-2}\Big)}$}&\\
&&\\
  \hline
   \multirow{4}{*}{${\bf \frac{1}{\sqrt{2}}\Big(\ket{d-2}\pm\ket{d-1}\Big)}$}&\multirow{2}{*}{${\bf \frac{1}{\sqrt{2}}\Big(\ket{d-3}\pm \ket{d-2}\Big)}$}&\multirow{2}{*}{${\bf \frac{d}{4}-1}$}\\
&&\\
&\multirow{2}{*}{${\bf \frac{1}{\sqrt{2}}\Big(\ket{d-1}\pm \ket{0}\Big)}$}&\multirow{2}{*}{${\bf \perp}$}\\
&&\\
  \hline
\end{tabular}
}
\caption*{(b)}
    \end{minipage} 
    \caption{ Key generation rule for (a) $d=4n$, and (b) $d=4n+2$. In the latter case, Key letters $0$ and $\frac{d}{4}-1$ are generated with probabilities $\frac{1}{2}$ in contrast to other key letters. The symbol $\perp$ corresponds to the rounds which are discarded.  This key generation rule is motivated by what is employed in B$92$ protocol \cite{PhysRevLett.68.3121} for unambiguous state discrimination.}
    \label{tab:rule_d}
\end{table}

\section{Semi-quantum key distribution (SQKD): a brief recap}
\label{SQKD_Boyer}
In this appendix, we briefly present the SQKDP proposed in \cite{Boyer07} (designated as SQKDP07), to make the paper self-contained. The SQKDP07  securely distributes a  key between a quantum participant (QP) ({\it viz.}, Alice) and a classical participant (CP) ({\it viz.}, Bob). A QP can prepare any state and perform measurements in any basis whereas a CP can prepare and measure only in the computational basis. Since the SQKDP has only one QP, it eases down the experimental implementation of the SQKDP07  \cite{massa2022experimental}. For reference, the steps of the SQKDP07  are  explicitly enumerated as follows:
\begin{enumerate}
    \item Alice randomly prepares one of the states from the two bases $B_0\equiv \{\ket{0}, \ket{1}\}, B_1\equiv  \{\ket{+}, \ket{-}\}$ with  equal probability and sends it to Bob. 
    \item Bob, upon receiving the state, exercises one of the two options with equal probabilities: (i) he measures the incoming state in the computational basis. He prepares the same state as the post-measurement state afresh and sends it back to Alice, (ii) He sends the incoming state back to Alice.  
    \item Alice measures each incoming state in the same basis in which she has prepared it. This constitutes one round.
    \item After a sufficient number of rounds, Bob reveals the rounds in which he has performed measurements  and Alice reveals the rounds in which she has sent the states in the computational basis, i.e., $\{\ket{0}, \ket{1}\}$. 
    \item Alice analyses the data of those rounds in which Bob has not performed measurements to check for the presence of an eavesdropper.
    \item The outcomes of those rounds, in which Bob has performed measurements and Alice has measured in the computational basis, constitute a key.
\end{enumerate}
In this way, a key is shared between Alice and Bob. 


\end{document}